\documentclass[apj,twocolappendix,numberedappendix]{emulateapj}

\usepackage{amsmath}
\usepackage{color}
\usepackage{hyperref}

\newcommand{\angstrom}{\textup{\AA}}

\usepackage{xcolor}

\usepackage[]{units}
\slugcomment{To be submitted to Astrophysical Journal Letters}

\begin{document}

\title{Gamma~rays from the quasar PKS~1441+25: story of an escape}
\author{
A.~U.~Abeysekara\altaffilmark{1},
S.~Archambault\altaffilmark{2},
A.~Archer\altaffilmark{3},
T.~Aune\altaffilmark{4},
A.~Barnacka\altaffilmark{5},
W.~Benbow\altaffilmark{6},
R.~Bird\altaffilmark{7},
J.~Biteau\altaffilmark{8,9,44},
J.~H.~Buckley\altaffilmark{3},
V.~Bugaev\altaffilmark{3}
% ...
}

\altaffiltext{1}{Department of Physics and Astronomy, University of Utah, Salt Lake City, UT 84112, USA}
\altaffiltext{2}{Physics Department, McGill University, Montreal, QC H3A 2T8, Canada}
\altaffiltext{3}{Department of Physics, Washington University, St.~Louis, MO 63130, USA}
\altaffiltext{4}{Department of Physics and Astronomy, University of California, Los Angeles, CA 90095, USA}
\altaffiltext{5}{Harvard-Smithsonian Center for Astrophysics, 60 Garden Street, Cambridge, MA 02138, USA}
\altaffiltext{6}{Fred Lawrence Whipple Observatory, Harvard-Smithsonian Center for Astrophysics, Amado, AZ 85645, USA}
\altaffiltext{7}{School of Physics, University College Dublin, Belfield, Dublin 4, Ireland}
\altaffiltext{8}{Santa Cruz Institute for Particle Physics and Department of Physics, University of California, Santa Cruz, CA 95064, USA}
\altaffiltext{9}{now at Insititut de Physique Nucl\'eaire d'Orsay (IPNO), CNRS-INP3, Univ.~Paris-Sud, Universit\'e Paris-Saclay, 91400 Orsay, France}
% ...
\altaffiltext{44}{jbiteau@ucsc.edu, biteau@ipno.in2p3.fr}
% ...

\shorttitle{Gamma~rays from PKS~1441+25}
\shortauthors{The VERITAS Collaboration {\it et al.}}

\begin{abstract}
Outbursts from gamma-ray quasars provide insights on the relativistic jets of active galactic nuclei and constraints on the diffuse radiation fields that fill the Universe. The detection of significant emission above $\unit[100]{GeV}$ from a distant quasar would show that some of the radiated gamma~rays escape pair-production interactions with low-energy photons, be it the extragalactic background light (EBL), or the radiation near the supermassive black hole lying at the jet's base. VERITAS detected gamma-ray emission up to $\sim\unit[200]{GeV}$ from PKS 1441+25 ($z=0.939$) during April 2015, a period of high activity across all wavelengths. This observation of PKS~1441+25 suggests that the emission region is located thousands of Schwarzschild radii away from the black hole. The gamma-ray detection also sets a stringent upper limit on the near-ultraviolet to near-infrared EBL intensity, suggesting that galaxy surveys have resolved most, if not all, of the sources of the EBL at these wavelengths.
\end{abstract}


\section{A new very-high-energy quasar}
\label{Sec:Intro}

The extragalactic gamma-ray sky is dominated by the emission of blazars, active galactic nuclei whose jets are pointed within a few degrees of Earth. About sixty blazars have been detected at very high energy (VHE; $E>\unit[100]{GeV}$),\footnote{http://tevcat.uchicago.edu/} with only four belonging to the class of flat-spectrum radio quasars (FSRQs): 3C~279 \citep[$z=0.536$,][]{2008Sci...320.1752M}, PKS~1510-089 \citep[$z=0.361$,][]{2013A&A...554A.107H}, PKS~1222+216 \citep[$z=0.432$,][]{2011ApJ...730L...8A}, and S3~0218+35 \citep[$z=0.944$,][]{2015arXiv150804580S}.

FSRQs are believed to host radiatively-efficient disks that enrich the environment of the supermassive black hole with ultraviolet-to-optical photons. This photon field, the reprocessed emission from the clouds of the broad line region (BLR), and the infrared radiation from the ``dusty torus" can all interact with gamma rays through pair production, preventing the escape of VHE radiation from the base of the jet \citep{2003APh....18..377D}.

VHE gamma~rays that do escape will face pair production on the extragalactic background light (EBL), which encompasses the ultraviolet-to-infrared emission of all stars and galaxies since the epoch of reionization ($z\lesssim 10$).\footnote{We adopt the concordance $\Lambda$CDM model ($h_0 = 0.7$, $\Omega_M=0.3$, $\Omega_\Lambda=0.7$).} Direct measurements of the EBL are prone to contamination from the bright local environment, while strict lower limits are derived from galaxy surveys, measuring the light emitted by known populations of sources \citep{2000MNRAS.312L...9M}.

VHE detections of high-redshift FSRQs constrain both the EBL and the gamma-ray emission regions in blazars. This letter reports the detection of VHE gamma~rays from the FSRQ PKS~1441+25 \citep[$z=0.939$,][]{2012ApJ...748...49S} by VERITAS. This source was observed in April and May 2015 following the VHE discovery by MAGIC \citep{2015ATel.7416....1M}, triggered by multiwavelength activity \citep{2015ATel.7402....1P} and a spectral hardening at high energies \citep[HE, $100\ {\rm MeV}<E<100\ {\rm GeV}$;][]{DaveThompsonPrivate}.

\section{Observations of PKS~1441+25}
\label{Sec:Obs}

\begin{figure}
\hspace{-0.4cm}
\includegraphics[width=0.50\textwidth]{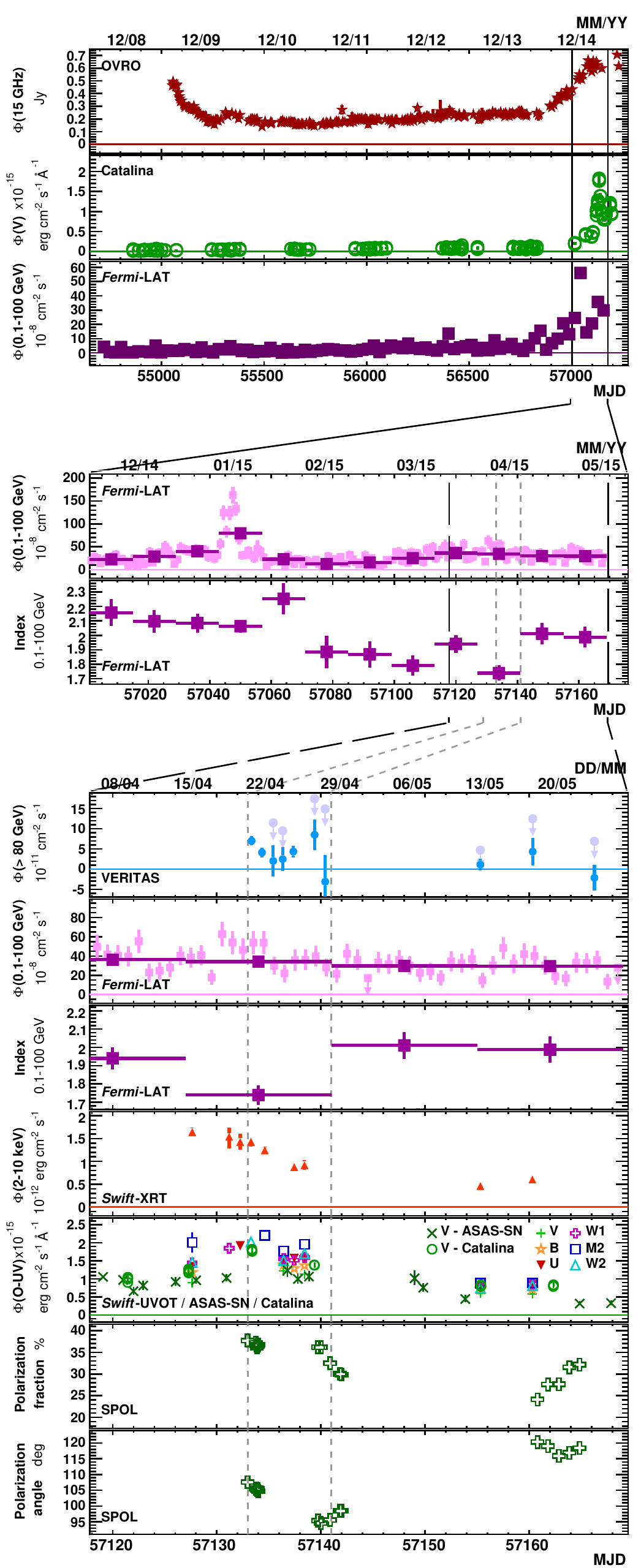}
\caption{{\it Top:} observations of PKS~1441+25 from 2008 to 2015. {\it Middle:} observations from December 2014 to May 2015. {\it Bottom:} observations in April and May. The gray dashed lines mark the period considered for the analyses in Sec.~\ref{Sec:SED} and \ref{Sec:Abs}.}
\label{fig_1}
\end{figure}

PKS~1441+25 was detected from 2015 April 21 (MJD~57133) to April 28 (MJD~57140) with VERITAS, an array of four imaging atmospheric Cherenkov telescopes located in southern Arizona \citep[][]{2011ICRC...12..137H}. VERITAS imaged gamma-ray induced showers from the source above $\unit[80]{GeV}$, enabling the detection of PKS~1441+25 (VER~J1443+250) at a position consistent with its radio location and at a significance of 7.7~standard~deviations ($\sigma$) during the $\unit[15.0]{h}$ exposure (2710 ON-source events, 13780 OFF-source events, OFF normalization of 1/6). Using a standard analysis with cuts optimized for low-energy showers \citep[][and references therein]{2014ApJ...785L..16A}, we measure an average flux of $\Phi(>\unit[80]{GeV}) = \unit[(5.0\pm0.7)\times10^{-11}]{cm^{-2}\ s^{-1}}$ with a photon index $\Gamma_{\rm VHE} = 5.3 \pm 0.5$ up to $\unit[200]{GeV}$,\footnote{The last spectral points at $140$ and $\unit[180]{GeV}$ are significant at the $2.4$ and $\unit[3.0]{\sigma}$ level, respectively} corresponding to an intrinsic index of $3.4\pm0.5$ after correction for the EBL \citep[][``fixed'']{2012MNRAS.422.3189G}. The day-by-day lightcurve is compatible with constant emission in that period ($\chi^2/ndf=7.4/6$), and fractional variability $F_{\rm var}<110\ \%$ at the $95\ \%$ confidence level \citep[][]{Vaughan}.\footnote{All flux estimates are used for variability constraints but we also show $\unit[99]{\%}$-confidence-level upper limits for points below $\unit[3]{\sigma}$ in Fig.~\ref{fig_1}} Subsequent observations in May (MJD~57155-57166, $\unit[3.8]{h}$ exposure) showed no significant excess  (660 ON-source events, 3770 OFF-source events, OFF normalization of 1/6), resulting in an upper limit of $\Phi(>\unit[80]{GeV}) < \unit[4.3\times10^{-11}]{cm^{-2}\ s^{-1}}$ at the $\unit[99]{\%}$ confidence level. These results have been cross checked with an independent calibration and analysis. Monte-Carlo simulations indicate systematic uncertainties on the VHE energy scale and photon index of $\unit[20]{\%}$ and $0.2$, respectively. The systematic uncertainty on the flux of this source is estimated to be $\unit[60]{\%}$, including the energy-scale uncertainty discussed in \citet[][]{2014ApJ...785L..16A}.   
 
The LAT pair-conversion telescope onboard the {\it Fermi} satellite has surveyed the whole sky in the HE band since August 2008 \citep{2009ApJ...697.1071A}. We analyzed the LAT data using the public science tools {\tt v10r0p5} (Pass-8) leaving free the parameters of sources from the 3FGL \citep[][]{2015ApJS..218...23A} within a region of interest of $\unit[10]{^\circ}$ radius and fixing them for sources $\unit[10-20]{^\circ}$ away. We reconstruct the spectrum of PKS~1441+25 between $\unit[100]{MeV}$ and $\unit[100]{GeV}$ in four-week (MJD~54705-57169, top panel in Fig.~\ref{fig_1}) and two-week (MJD~57001-57169, middle panel) bins assuming a power-law model with a free normalization and photon index (purple points), as well as in one-day bins (pink points) fixing the photon index to its best-fit average value in MJD~57001-57169, $\Gamma_{\rm HE} = 1.97 \pm 0.02$, slightly harder than in the 3FGL, $2.13\pm0.07$. The source was in a high state during MJD~57001-57169, with an integrated $\unit[100]{MeV}-\unit[100]{GeV}$ flux that is one to two orders of magnitude above the 3FGL value, $\unit[(1.3\pm0.1)\times10^{-8}]{cm^{-2}\ s^{-1}}$. During the period contemporaneous with the VERITAS detection (MJD~57133-57140), the source shows a flux of $\unit[(34\pm4)\times10^{-8}]{cm^{-2}\ s^{-1}}$ and a hard index of $1.75\pm0.08$. Although a power-law model is used for robustness in the lightcurve determination, the spectrum shows a hint of curvature, with a log parabola preferred over a power law by $\unit[3.2]{\sigma}$ (see Fig.~\ref{fig_2}). The curvature is resilient to changes in the analysis and the temporal window, and fits in smaller energy ranges confirm the hint.

X-ray observations with {\it NuSTAR} and {\it Swift} were triggered following the VHE detection. {\it NuSTAR}, a hard-X-ray instrument sensitive to $\unit[3-79]{keV}$ photons \citep{2013ApJ...770..103H}, observed the source on MJD~57137 for an exposure of $\unit[38.2]{ks}$. The data were reduced using the {\tt NuSTARDAS} software v1.3.1. {\it Swift}-XRT \citep{2004ApJ...611.1005G} observed PKS~1441+25 between $0.3$ and $\unit[10]{keV}$ in June 2010 (MJD~55359), in January 2015 (MJD~57028 \& 57050), in April 2015 (MJD~57127-57138), and in May 2015 (MJD~57155 \& 57160). Data taken in photon-counting mode were calibrated and cleaned with {\tt xrtpipeline} using CALDB 20140120 v.014. ON-source and background events were selected within regions of 20-pixel ($\sim\unit[46]{arcsec}$) and 40-pixel radius, respectively. The XRT and {\it NuSTAR} spectral analyses were performed with {\tt XPSEC} v12.8.2, requiring at least 20 counts per bin. The {\it NuSTAR} spectrum is matched by a power law with a photon index of $2.30 \pm 0.10$ and an integrated $\unit[3-30]{keV}$ flux of $\unit[(1.25\pm0.09)\times10^{-12}]{erg\ cm^{-2}\ s^{-1}}$. No intranight variability is detected. {\it Swift}-XRT did not significantly detect the source in June 2010, but the 2015 observations reveal a power-law spectrum with no detectable spectral variability and an average $0.3-\unit[10]{keV}$ photon index of $2.35 \pm 0.24$, using an absorbed model with a hydrogen column density of $\unit[3.19\times10^{20}]{cm^{-2}}$ \citep{2005A&A...440..775K}. Significant flux variations are detected in {the period contemporaneous with VERITAS observations ($\chi^2/ndf=25.9/3$, $F_{\rm var}=\unit[22.6\pm0.9]{\%}$), with a flux-halving time of $\unit[13.9\pm1.4]{days}$ based on an exponential fit to the data in MJD~57127-57155 ($\chi^2/ndf=8.6/6$).

\begin{figure*}
\hspace{-0.2cm}
\includegraphics[width=1.02\textwidth]{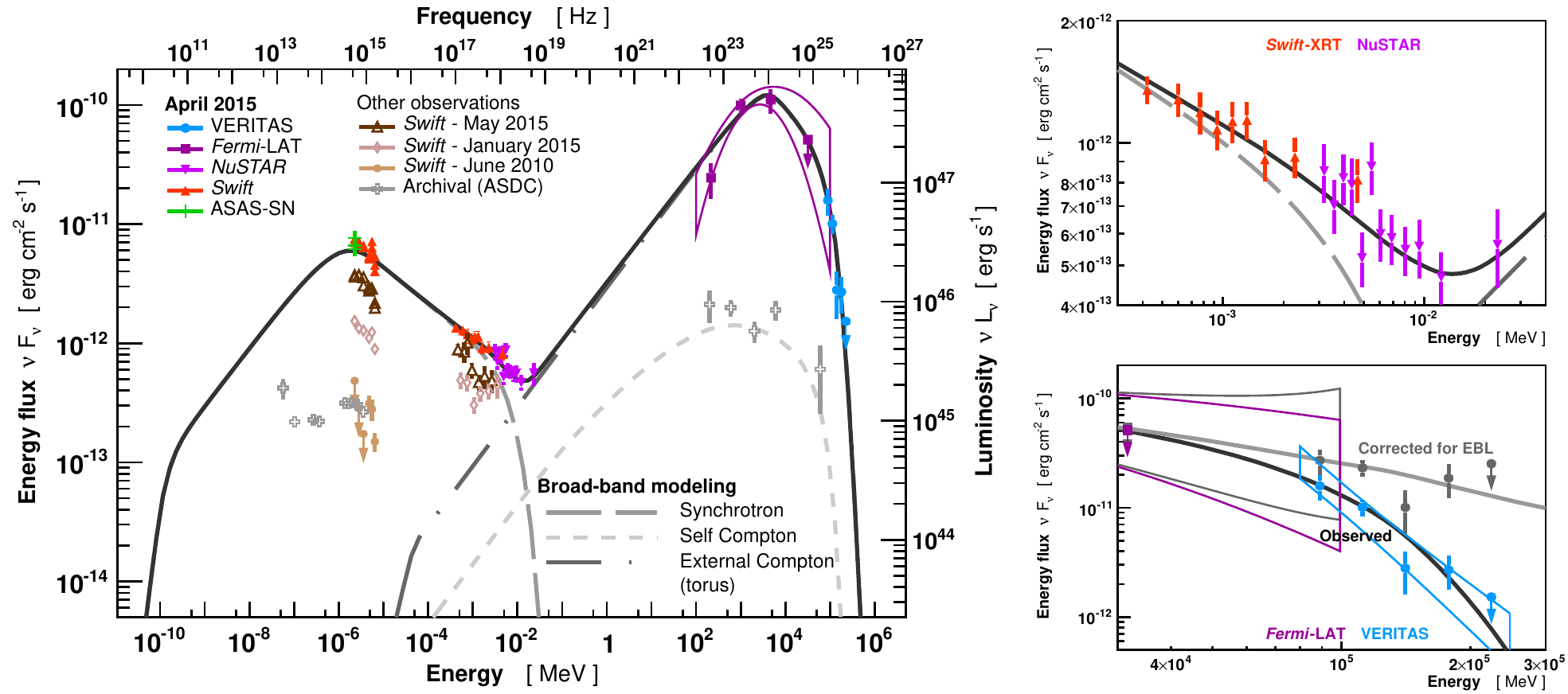}
\caption{Multiwavelength emission of PKS~1441+25. Side panels show the X-ray (top) and gamma-ray emission (bottom) in April 2015 (MJD~57133-57140). The various exposures and the model are discussed in Sec.~\ref{Sec:Obs} and \ref{Sec:SED}, respectively.}
\label{fig_2}
\end{figure*}

Simultaneously with the XRT observations, {\it Swift}-UVOT \citep{RomingUVOT} took photometric snapshots of PKS~1441+25 in six optical-to-ultraviolet filters. Flux densities were extracted using  \textit{uvotmaghist} and circular ON-source and background regions of $\unit[5]{arcsec}$ and $\unit[15]{arcsec}$ radius, respectively. PKS~1441+25 has been observed in the V-band since January 2012 within the All-Sky Automated Survey for Supernovae \citep[ASAS-SN;][]{2014ApJ...788...48S}, using the quadruple 14-cm ``Brutus" telescope in Hawaii. The fluxes from both experiments in Fig.~\ref{fig_2} are dereddenned using $E(B-V)=0.043$, consistent with the column density used for the XRT analysis \citep{1974ApJ...187..243J}. The 0.68-m Catalina Schmidt Telescope (AZ) has also performed long-term unfiltered optical observations of PKS 1441+25 since 2005 within the Catalina Real-Time Transient Survey \citep[CRTS,][]{2009ApJ...696..870D}. Observed magnitudes were converted into the V-band using the empirical method described in \cite{2013ApJ...763...32D}. The SPOL spectropolarimeter \citep{1992ApJ...398L..57S} has monitored the linear optical polarization of PKS~1441+25 in $\unit[5000-7000]{\angstrom}$, with observations at the 1.54-m Kuiper Telescope, at the 6.5-m MMT, and at the Steward Observatory 2.3-m Bok Telescope (AZ). The source shows a high degree of polarization, with values ranging from $\unit[37.7\pm0.1]{\%}$ to $\unit[36.2\pm0.1]{\%}$ between MJD~57133 and MJD~57140.

The OVRO 40-m telescope \citep{2011ApJS..194...29R} has monitored PKS~1441+25 at $\unit[15]{GHz}$ since late 2009. A $\unit[15]{GHz}$ VLBA image obtained by the MOJAVE program \citep{2009AJ....138.1874L} on 2014 March 30 (MJD~56381) shows a compact core and a bright, linearly polarized jet feature located $\unit[1.2]{milliarcsec}$ downstream, at position angle $\unit[-68]{^\circ}$. Both features have relatively high fractional polarization ($\sim\unit[10]{\%}$), and electric vectors aligned with the jet direction, at an angle of $\unit[102]{^\circ}$ similar to that measured by SPOL, indicating a well-ordered transverse magnetic field. The fractional polarization level of the core feature is among the highest seen in the MOJAVE program \citep{2011ApJ...742...27L}.

The 2008-2015 observations of PKS~1441+25 shown in Fig.~\ref{fig_1} reveal a brightening of the source in the radio, optical, and HE bands starting around MJD~56900. A simple Pearson test \citep[see caveats in][]{2014MNRAS.445..437M} applied to the radio and HE long-term lightcurves shows a correlation coefficient $r=0.75\pm0.02$, differing from zero by $\unit[5.4]{\sigma}$ based on the $r$-distribution of shuffled lightcurve points. Similarly, the analysis of the optical and HE lightcurves yields $r=0.89\pm0.02$, differing from zero by $\unit[4.8]{\sigma}$. The discrete correlation functions display broad, zero-centered peaks with widths of $\sim\unit[100]{days}$, indicating no significant time lags beyond this time scale. During the period marked by gray dashed lines in Fig.~\ref{fig_1}, observations on daily timescales from optical wavelengths to X-rays reveal fractional flux variations smaller than $\unit[25]{\%}$, compatible with the upper limits set by {\it Fermi}-LAT and VERITAS ($\unit[30]{\%}$ and $\unit[110]{\%}$ at the $\unit[95]{\%}$ confidence level, respectively). Such flux variations are small with respect to the four orders of magnitude spanned in $\nu F_{\nu}$, enabling the construction of a quasi-contemporaneous spectral energy distribution in Sec.~\ref{Sec:SED}.

\section{Emission Scenario}
\label{Sec:SED}

The spectral energy distribution, with the X-ray-to-VHE data averaged over the active phase in April 2015 (MJD~57133-57140), is shown in Fig~\ref{fig_2}. The optical-to-X-ray spectrum is well described by a power law with photon index $\Gamma=2.29\pm 0.01$ from $\unit[2]{eV}$ to $\unit[30]{keV}$, including a $\unit[10]{\%}$ intrinsic scatter in the fit procedure that accounts for the small-amplitude optical-to-UV variability. 
This spectrum suggests a single synchrotron component peaking below $\unit[2]{eV}\sim\unit[5\times10^{14}]{Hz}$, created by an electron population of index $p = 2\Gamma-1 \sim 3.58 \pm 0.02$. As expected in FSRQs \citep{1998MNRAS.299..433F}, the emission of PKS~1441+25 is dominated by the gamma-ray component, well-described by a single component peaking at $\unit[3.3_{-1.1}^{+1.8}]{GeV}$.}

The detection of gamma~rays up to $\unit[200]{GeV}$, about $\unit[400]{GeV}$ in the galaxy's frame, suggests that the emitting region is located beyond the BLR, or else pair production would suppress any VHE flux even for a flat BLR geometry \citep{2012arXiv1209.2291T}. The elevated radio state, correlated with the optical and HE brightening, also suggests synchrotron emission outside of the BLR where synchrotron self-absorption is smaller. The hypothesis of large-scale emission is strengthened by the week-long duration of the optical-to-gamma-ray flare. This behavior contrasts with other observations of bright FSRQs, displaying different flux variations at different wavelengths \citep[e.g.][]{2010Natur.463..919A}, more in line with multi-component scenarios. The flare of PKS~1441+25 appears to be one of the few events whose detailed temporal and spectral multiwavelength features are consistent with the emission of a single component beyond the BLR.

The BLR size can be derived using the estimated black-hole mass, $M_{\rm BH} = 10^{7.83 \pm 0.13} M_\odot$ \citep{2012ApJ...748...49S}, assuming $r_{\rm BLR} \simeq \unit[10^{17}]{cm} \times \sqrt{L_{\rm disk}/\unit[10^{45}]{erg\ s^{-1}}}$ \citep{2007ApJ...659..997K} and an accretion disk luminosity that is a fraction $\eta = \unit[10]{\%}$ of the Eddington luminosity. Alternatively, $L_{\rm disk}$ can be estimated from the BLR luminosity as $L_{\rm disk} \simeq 10\times L_{BLR}$, with $L_{BLR}=\unit[10^{44.3}]{erg\ s^{-1}}$ \citep{2014MNRAS.441.3375X}. Both estimates yield $r_{\rm BLR} \simeq \unit[0.03]{pc}$, setting a lower limit on the distance between the black hole and the emitting region of $r\gtrsim5,000$ Schwarzschild radii. 

We show in Fig.~\ref{fig_2} that a stationary model can reproduce the data, using the numerical code of \cite{2013ApJ...771L...4C} and the EBL model of \cite{2012MNRAS.422.3189G}. Non-stationary Klein-Nishina effects on electron cooling could be important for this source \citep{2005MNRAS.363..954M}. Nonetheless, the straight power law observed from optical to X-rays suggests that the upscattering electrons see photon energies that are low enough to stay out of the Klein-Nishina regime. This again indicates that the emission is outside of the BLR. We parametrize the electron-positron population by a fixed broken power law with indices $p_1=2$ and $p_2=3.58$ between $\gamma_{\rm min}=1$ and $\gamma_{\rm max}=7\times10^5$, with a break at $\gamma_{\rm break}=1.2\times10^4$ (jet frame). The particle energy density is on the order of the magnetic energy density, $u_e/u_B = 1.5$, with a total luminosity of $\unit[6\times10^{45}]{erg\ s^{-1}}$, and an infrared photon energy density of $1.4\times\unit[10^{-5}]{erg\ cm^{-3}}$ at $T=\unit[10^3]{K}$ (galaxy frame), characteristic of thermal radiation from the torus. We broke the degeneracies among the parameters of the model by requiring that the minimum variability timescale in the observer frame, $t_{\rm var} = (1+z)/\delta \times R/c$, be comparable to the flux-halving timescale of about two weeks observed in X rays, where we have the best statistics to probe variability. The data are then modeled with an emitting region of radius $R=\unit[4\times10^{17}]{cm}\sim\unit[0.1]{pc}$ and Doppler factor $\delta \sim 18$. The magnetic field, $B\sim\unit[80]{mG}$, is tangled within the emitting frame, but compressed transversely to the motion within the observer's frame, which would explain the high optical-polarization degree, $PD$. Following \cite{2014ApJ...784..141S}, the ratio of $PD$ over the maximum theoretical polarization $\Pi_S = (p+1)/(p+7/3)$ constrains the angle $\theta$ at which the emitting region is viewed. We find that the Doppler factor and the geometry of the system are well reproduced by a jet Lorentz factor $\Gamma_{\rm jet} \sim 12$ and $\theta \sim \unit[2.6]{^\circ}$, which is within the jet opening angle, $\theta_{\rm jet} = 1/\Gamma_{\rm jet} \sim \unit[4.8]{^\circ}$. 

The location of the emission can be roughly estimated assuming that the whole cross-section of the jet contributes to the radiation \citep{2010MNRAS.405L..94T}, as $r \sim R / \tan\theta_{\rm jet} \sim \unit[1.5]{pc}$, or 200,000 Schwarzschild radii. The region is not expected to be much more compact than in this model, as no fast large-amplitude variability is seen from optical to X-ray wavelengths, despite the statistics being sufficient to detect doubling timescales as short as days. The region could still be further away from the black hole if its size was only fraction of the jet cross-section.    

The model parameters are similar to those obtained by \cite{2008Sci...320.1752M}, \cite{2008AIPC.1085..427B}, \cite{2011A&A...534A..86T}, and \cite{2014A&A...567A.113B} for other FSRQs, but a remarkably high break energy in the electron spectrum is needed to explain the optical-to-X-ray synchrotron emission.  The break in the electron distribution is consistent with radiative cooling, but is pushed to higher energies due to the magnetic field being two to ten times lower than that inferred for other VHE FSRQs. The magnetic field is also in equipartition with the particle population, minimizing the energy budget required to produce the synchrotron emission. Finally, the jet Lorentz factor is two to four times smaller than that required for other VHE FSRQs, highlighting again the reasonable energetics of this scenario.

\section{Extragalactic Background Light}
\label{Sec:Abs}

The redshift of PKS~1441+25, $z=0.939$, and its detection up to $\unit[200]{GeV}$ provide an exceptional opportunity to study the EBL. 

\begin{figure}
\hspace{-0.15cm}
\includegraphics[width=0.53\textwidth]{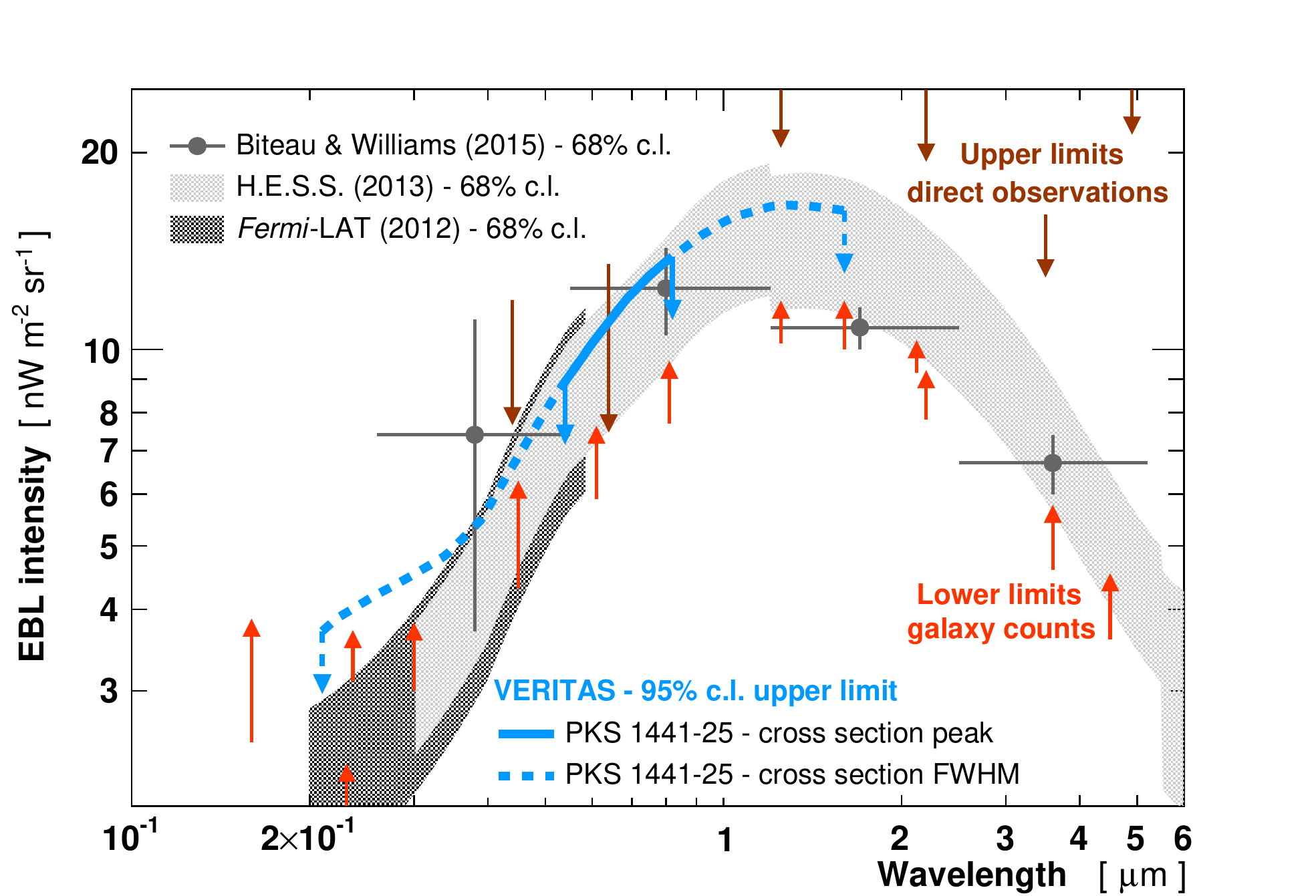}
\caption{Near-ultraviolet to near-infrared spectrum of the EBL. The upper limit from this work is shown in blue, in regions corresponding to the peak and FWHM of the cross section ($1<\tau<2$).}
\label{fig_3}
\end{figure}

Gamma~rays interact with EBL photons through pair production, yielding an observed spectrum that is softer than the intrinsic spectrum. Imposing a maximum intrinsic VHE hardness can then constrain the EBL intensity \citep{2006Natur.440.1018A}. In a scenario where the HE and VHE energy photons originate from the same component \citep[but see][for emission within the BLR]{2014ApJ...794....8S}, {the unattenuated HE observations set an upper limit on the intrinsic hardness. Based on Sec.~\ref{Sec:SED}, we neglect any additional emission component and fit, as in \cite{2015arXiv150204166B}, an absorbed power law with free EBL normalization, $\alpha$, to the VERITAS spectrum. We minimize 

\begin{align}
\label{Eq:chi2}
\chi^2 =& \sum_{i=1..n} \frac{(\phi_i - \phi_0\times{\rm e}^{-\Gamma\log(E_i/E_0)-\alpha\tau(E_i)})^2}{\sigma_{\phi_i}^2}\nonumber\\
&+\Theta(\Gamma_{\rm LAT}-\Gamma)\times\frac{(\Gamma_{\rm LAT}-\Gamma)^2}{\sigma_{\Gamma}^2}, 
\end{align}
where $\alpha$, $\phi_0$, and $\Gamma$ are free parameters, $E_0$ is fixed to $\unit[120]{GeV}$, $\{\phi_i\}_{i=1..n}$ are the fluxes measured by VERITAS at energies $\{E_i\}_{i=1..n}$, and $\Theta$ is the Heaviside function. The {\it Fermi}-LAT log-parabolic spectrum in Fig.~\ref{fig_2} shows a photon index $\Gamma_{\rm LAT} = 2.76\pm0.43$ at $\unit[30]{GeV}$, where the absorption by the EBL is smaller than $\unit[5]{\%}$ \citep[][]{2008A&A...487..837F,2011MNRAS.410.2556D,2012MNRAS.422.3189G}. We account for the systematic uncertainty on the VERITAS photon index, $0.20$, by imposing a maximum hardness with uncertainty $\sigma_\Gamma = 0.43 \oplus 0.20  = 0.47$, where $\oplus$ indicates a quadratic sum. We finally marginalize the equivalent likelihood, $\exp(-\chi^2/2)$, over the VERITAS energy scale. The logarithm of the latter is assumed to be Gaussian, with zero mean and width $0.2$, corresponding to a $\unit[20]{\%}$ systematic uncertainty. Equation~\ref{Eq:chi2} allows for an intrinsic VHE spectrum that is softer, but not harder, than the HE spectrum. This yields an EBL normalization that is consistent with $\alpha=0$ and constrained to $\alpha<1.5$ at the $\unit[95]{\%}$ confidence level for the model of \cite{2012MNRAS.422.3189G}. This result is almost independent of the choice of model \cite[][]{2008A&A...487..837F,2011MNRAS.410.2556D}. 

Considering both the peak and full width at half maximum (FWHM) of the cross section integrated along the line-of-sight \citep[as in][with various evolutions tested]{2015arXiv150204166B}, the VERITAS observations constrain the near-ultraviolet to near-infrared EBL. Our constraint shown in Fig.~\ref{fig_3}} is compatible and competitive below $\unit[1]{\mu m}$ with other state-of-the-art gamma-ray measurements from \cite{2012Sci...338.1190A}, \cite{2013A&A...550A...4H}, and \cite{2015arXiv150204166B}. Although $\alpha$ is compatible with zero, there is no significant tension with local constraints \citep[see][brown and orange arrows in Fig.~\ref{fig_3}]{2015arXiv150204166B}, since the differences are only $\unit[1.5]{\sigma}$ and $\unit[1.7]{\sigma}$ in the peak and FWHM regions, respectively.

\newpage

\section{Discussion}

The low energy threshold of VERITAS enabled the detection above $\unit[80]{GeV}$ of one of the most distant VHE gamma-ray sources ($z=0.939$), in a redshift range previously accessible only to space-borne gamma-ray observatories. We obtain stringent constraints on the EBL intensity below $\unit[1]{\mu m}$ and conclude that galaxy surveys have resolved most, if not all, of the sources of the EBL in this region. This provides an excellent baseline for studies above $\unit[1]{\mu m}$ where the redshifted ultraviolet emission of primordial stars could be detected \citep{2005ApJ...635..784D,2015arXiv150204166B}. 

The VHE detection of the highly-polarized source PKS~1441+25 is contemporaneous with a period of hard HE emission and of enhanced flux at all wavelengths. The correlation between the radio, optical, and HE lightcurves, unusual for this class of sources \citep{2014MNRAS.445..428M}, together with slow multiwavelength variability, suggest that the multi-band flare was produced by a single region located $\sim10^4-10^5$ Schwarzschild radii away from the black hole, which is consistent with the VHE-gamma-ray escape condition.

PKS~1441+25 is by far the dimmest HE emitter of all VHE-detected FSRQs listed in the 3FGL catalog. While HE activity remains a prime trigger of VHE observations, searches for new VHE-emitting quasars could also factor in radio-to-optical brightening and synchrotron-dominated X-ray emission, as reported for PKS~1441+25. These criteria will be of particular interest if applied to distant FSRQs, possibly opening a new observational window on the jets of blazars and on the transformation of the Universe's light content with cosmic time.

\acknowledgments
This research is supported by grants from the U.S. Department of Energy Office of Science, the U.S. National Science Foundation and the Smithsonian Institution, and by NSERC in Canada, with additional support from NASA Swift GI grant NNX15AR38G. We acknowledge the excellent work of the technical support staff at the Fred Lawrence Whipple Observatory and at the collaborating institutions in the construction and operation of the instrument. The VERITAS Collaboration is grateful to Trevor Weekes for his seminal contributions and leadership in the field of VHE gamma-ray astrophysics, which made this study possible. 

ASAS-SN thanks LCOGT, NSF, Mt. Cuba Astronomical Foundation, OSU/CCAPP and MAS/Chile for their support.

The observations at Steward Observatory are funded through NASA Fermi GI grant NNX12AO93G.

CRTS is supported by the NSF grants AST-1313422 and AST-1413600.

The OVRO 40-m monitoring program is supported in part by NASA grants NNX08AW31G and NNX11A043G, and NSF grants AST-0808050 and AST-1109911. 

The National Radio Astronomy Observatory is a facility of NSF operated under cooperative agreement by Associated Universities, Inc.

This research has made use of data from the MOJAVE database that is maintained by the MOJAVE team \citep{2009AJ....138.1874L}.

\end{document}